\documentclass[12pt]{article}
\usepackage{amsmath,amssymb}
\usepackage{graphicx}
\usepackage{subfigure}
\usepackage[pdftex]{hyperref}
\DeclareGraphicsExtensions{.eps,.bmp,.wmf,.jpg,.pdf}
\numberwithin{equation}{section}
\def\be{\begin{equation}}
\def\ee{\end{equation}}

\def\bea{\begin{eqnarray}}
\def\eea{\end{eqnarray}}

\title{Reconstructing the f(R) gravity from the holographic principle}
\author{L.N. Granda\thanks{ngranda@univalle.edu.co} \\
Department of Physics, Universidad del Valle\\ A.A. 25360, Cali,
Colombia} 
\date{}
\begin{document}
\maketitle

\begin{abstract}
\noindent An holographic f(R) gravity model of dark energy is proposed. The correspondence between the f(R) geometrical effective energy density with the holographic density, allows the reconstruction of the f(R) gravity in flat FRW background in the Einstein frame. The proposed infrared cut-off for the holographic energy density depends on two parameters which are fit using the luminosity versus redshift data, allowing a suitable reconstruction in two representative cases of the EoS parameter: for $\omega>-1$ and $\omega<-1$.
\end{abstract}
\section{Introduction}
\noindent Many astrophysical data, such as observations of large scale structure \cite{tegmark}, searches for type Ia supernovae \cite{riess}, and measurements of the cosmic microwave background anisotropy \cite{spergel}, all indicate that the expansion of the universe is undergoing cosmic acceleration at the present time, due to some
kind of negative-pressure form of matter known as dark energy (for a review see\cite{copeland},\cite{padmana}). Although the cosmological observations suggest that dark energy component is about $2/3$ of the total content of the universe, the nature of the dark energy as well as its cosmological origin remain unknown at present. Among other approaches related with a variety of scalar fields (see \cite{copeland}), a very promising approach to dark energy is related with the modified theories of gravity known as $f(R)$ gravity, in which dark energy emerges from the modification of geometry \cite{sergei1},\cite{sergei2},\cite{sergei3},\cite{carroll},\cite{capozziello}. Modified gravity gives a natural unification of the early-time inflation and late-time acceleration thanks to different role of gravitational terms relevant at small and at large curvature and may naturally describe the transition from deceleration to acceleration in the cosmological dynamics. Among others, modified gravity is one of the areas of the gravitational science in which S. Odintsov has contributed to develop thanks to a great number of pioneer investigations. Thus, the connection of modified gravity theories with M-theory was suggested in \cite{sergei1} and some explicit asymptotic examples were shown. In \cite{sergei2},\cite{sergei3},\cite{sergei4},\cite{sergei41},\cite{sergeinoj}, a modified gravity model with positive and negative powers of curvature and $\ln R$ terms were proposed, to describe different unification scenarios of early time inflation with the late-time cosmic acceleration. 
This unification was also studied with a Gauss-Bonnet invariant dependent function as proposed in \cite{sergei5}, \cite{sergei6}. The one-loop quantization approach to $f(R)$ gravity was investigated in \cite{sergei7}. The asymptotic behavior of a wide variety of $f(R)$ models, for different cosmological epochs in the FRW background has been presented in \cite{sergei8} and a consistency with experimental tests has been studied in \cite{sergei8},\cite{sergei9},\cite{sergei10},\cite{sergei11}. A comprehensive work on reconstruction of different modified $f(R)$ models, has been done trough the works \cite{sergei12},\cite{sergei13},\cite{sergei14},\cite{sergei15}, \cite{sergei16}, where it was explicitly demonstrated which versions of above theories may be reconstructed from the known universe expansion history (see also \cite{capozziello}, \cite{cardone}, \cite{xing} for $f(R)$ reconstruction). Other generalizations like a non-local model of f(R) gravity, non-minimal coupling of f(R) gravity with matter, with Maxwell theory and non-minimal Yang-Mills coupling were discussed in \cite{sergei17,sergei18,sergei19,sergei20} respectively. The structure of future singularities in models of $f(R)$ gravity consistent with local tests, have been worked in  \cite{sergei21}, \cite{sergei22}, and the constraining of some $f(R)$ models according to VIRGO, LIGO and LISA experiments have been investigated in \cite{sergei23}.\\
Another way to the solution of the dark energy problem within the fundamental theory framework, is related with some facts of the quantum gravity theory, known as the holographic principle. Applied to the dark energy issue, this principle establishes an infrared cut-off for the so called holographic energy density, related with cosmological scales (discussion and references are given in \cite{li},\cite{granda}). 
An infrared cut-off given by the future event horizon was proposed in \cite{li} and a generalized holographic infrared cut-off depending on local and non-local quantities was proposed in \cite{sergei24}.
Viewing the modified $f(R)$ gravity models as an effective description of the
underlying theory of dark energy, and considering the holographic vacuum energy scenario as pointing in the direction of the underlying theory of dark energy, it is interesting to study how the $f(R)$ gravity can describe the holographic energy density as an effective theory. The reconstruction of $f(R)$ theories under different conditions has been presented in \cite{sergei12,sergei13,sergei14,sergei15,sergei16,cardone,xing} and the holographic reconstruction of $f(R)$ gravity have been discussed in \cite{xing} with the infrared cut-off given by the future event horizon. In this contribution, an holographic $f(R)$ reconstruction using the new infrared cut-off given in \cite{granda,granda1}, is presented.
 
\section{The holographic model}
Let us start with the holographic dark energy density given by \cite{granda,granda1}
\begin{equation}\label{eq2}
\rho_{\Lambda}=3M_p^2\left(\alpha H^2+\beta \dot{H}\right)
\end{equation}
where $H=\dot{a}/a$ is the Hubble parameter and $\alpha, \beta$ are constants which must satisfy the restrictions imposed by the current observational data. This kind of density may appear as the simplest case of more general $f(H,\dot{H})$ holographic density in the FRW background. Note that this proposal avoids conflict with the causality. Here the constants $\alpha$ and $\beta$ will be fixed according to the luminosity versus redshift observational data as given in \cite{padmanabhan}, based on \cite{riess1} data.
As we will focus on late time reconstruction, we consider that the holographic dark energy density gives the dominant contribution to the Friedmann equation and neglect the contribution from matter and radiation, thus the Friedmann equation becomes simpler
\begin{equation}\label{eq3}
H^2=\frac{1}{3M_p^2} \rho_{\Lambda}=\alpha H^2+\beta \dot{H}
\end{equation}
which gives a power-law solution \cite{granda1}
\begin{equation}\label{eq4}
H=\frac{\beta}{\alpha-1}\frac{1}{t}
\end{equation}
with constant equation of state parameter given by
\begin{equation}\label{eq5}
\omega_{\Lambda}=-1+\frac{2}{3}\frac{\alpha-1}{\beta}
\end{equation}
In terms of the redshift parameter we can write $H$ in Eq. (\ref{eq4}) as follows
\begin{equation}\label{eq6}
H=H_0\left(1+z\right)^{(\alpha-1)/\beta}
\end{equation}
where $H_0=\frac{\beta}{\alpha-1}\frac{1}{t_0}$ and $t_0$ is the present time ($z=0$).
In the flat FRW background the luminosity distance $L_d$ can be written as \cite{copeland}
\begin{equation}\label{eq7}
d_L=(1+z)\int^{z}_{0}\frac{dz}{H(z)}
\end{equation}
where we used the light speed $c=1$. With $H(z)$ given by (\ref{eq6}) the luminosity distance takes the explicit form from (\ref{eq7})
\begin{equation}\label{eq8}
d_L(z)=\frac{1+z}{H_0}\frac{\beta}{\beta-\alpha+1}\left[\left(1+z\right)^{(\beta-\alpha+1)/\beta}-1\right]
\end{equation}
based on this result we can plot the luminosity distance $L_d$ in terms of the redshift, and tune the parameters $\alpha$ and $\beta$ in such a way that
may fit the data reconstructed plot as shown in \cite{padmana} and \cite{padmanabhan}. The plot for three representative values of the $\alpha$ and $\beta$ parameters is given in Fig. 1
\begin{center}
\includegraphics [scale=0.7]{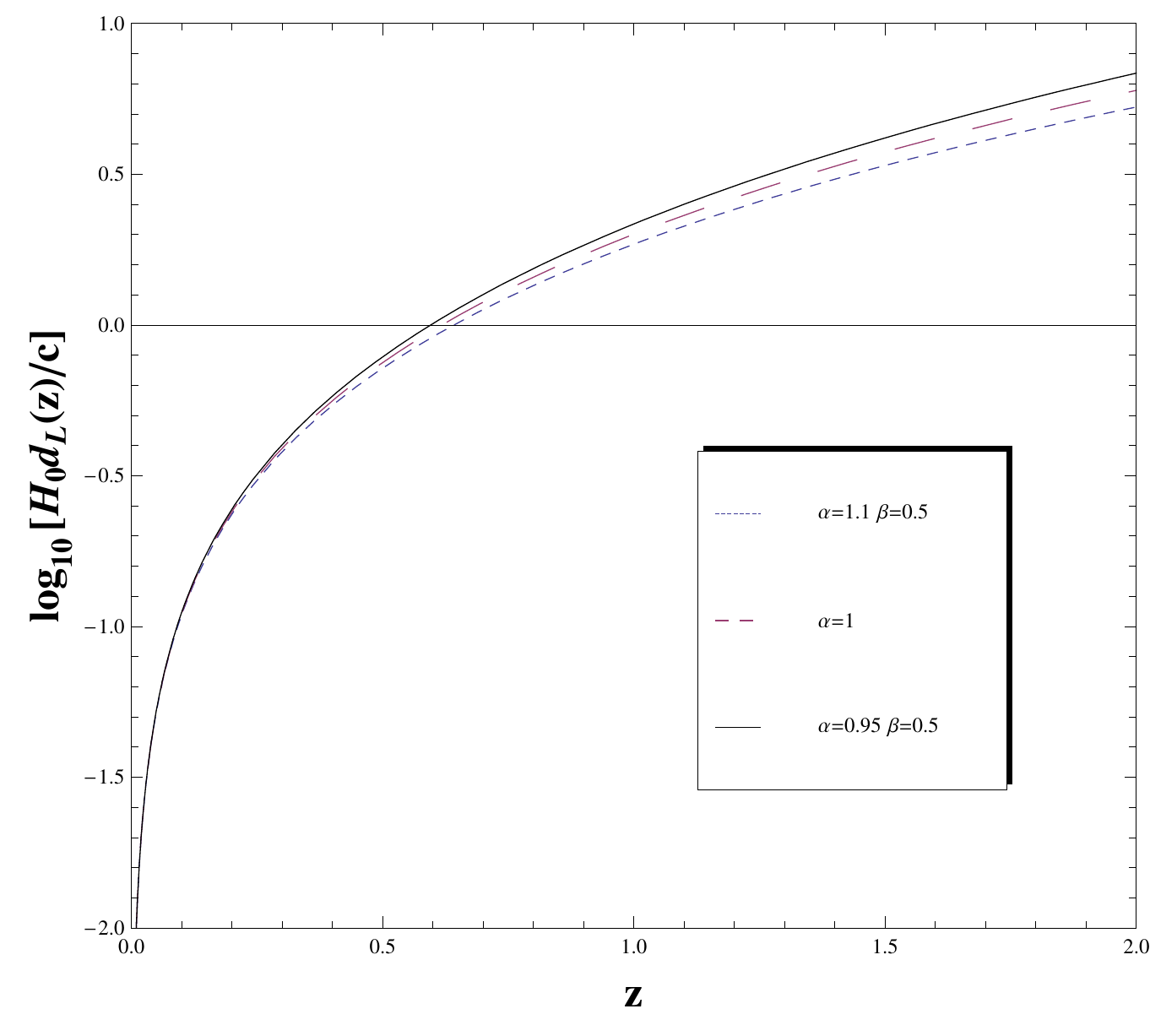}
\end{center}
\begin{center}
\it{Figure 1: The luminosity distance  $\log_{10}(H_0d_L(z))$ versus redshift. The three curves show $H_0d_L$ in logarithmic scale for $\Omega_{m0}=0$, $\Omega_{\Lambda0}=1$, $\alpha=1.1, 1, 0.95$ and $\beta=0.5$}. The $\alpha=1$ curve corresponds to the de Sitter solution.
\end{center}

\section{Late time reconstruction of f(R)}
Let's start with the action for f(R) gravity with matter
\be\label{eq9}
S=\int d^4x\sqrt{-g}\left[\frac{1}{2\kappa^2}f(R) + {\cal L}_m\right]
\ee
where $f(R)$ is a generic function of the Ricci scalar curvature $R$, and ${\cal L}_m$ is the matter fields Lagrangian density. Variation with respect to the metric and setting aside the complications with the variations of $f(R)$-type actions leads to the field equation \cite{sergei22},\cite{sotiriou}
\be\label{eq10}
f'(R)R_{\mu\nu}-\frac{1}{2}f(R)g_{\mu\nu}-\left(\nabla_{\mu}\nabla_{\nu}-g_{\mu\nu}\Box\right)f'(R)=\kappa^2T^{(m)}_{\mu\nu}
\ee
where $\nabla_{\mu}$ is the covariant derivative, prime denotes the derivative with respect to $R$ and $T^{(m)}_{\mu\nu}$ is the stress-energy tensor for the standard matter source. Interpreting this field equations in the form of Einstein equations with an effective stress-energy tensor composed of curvature terms moved to the right hand side, Eq. (\ref{eq10}) can be written as (hereafter we will take $\kappa^2=1$)
\be\label{eq11}
R_{\mu\nu}-\frac{1}{2}g_{\mu\nu}R=\frac{T^{(m)}_{\mu\nu}}{f'(R)}+T^{(R)}_{\mu\nu}
\ee
with $T^{(R)}_{\mu\nu}$ given by
\be\label{eq12}
T^{(R)}_{\mu\nu}=\frac{1}{f'(R)}\left[\frac{1}{2}\left(f(R)-Rf'(R)\right)g_{\mu\nu}+\nabla_{\mu}\nabla_{\nu}f'(R)-g_{\mu\nu}\Box f'(R)\right]
\ee
where $T^{(R)}_{\mu\nu}$ may be defined as an effective stress-energy tensor for some kind of curvature fluid. This allows to put Eq. (\ref{eq12}) in the form of a perfect fluid energy-momentum tensor, which will turn out to be useful in our reconstruction scheme. Note also that in this case the matter term becomes coupled to the geometry through $f'(R)$.\\
Assuming flat FRW background, the modified Friedmann equations take the form 
\be\label{eq13}
H^2=\frac{1}{3f'(R)}\left[\rho_m+\frac{Rf'(R)-f(R)}{2}-3H\dot{R}f''(R)\right]
\ee
in this way we can define an effective curvature-density
\be\label{eq14}
\rho_R=\frac{1}{f'(R)}\left[\frac{Rf'(R)-f(R)}{2}-3H\dot{R}f''(R)\right]
\ee
The space-space components of (\ref{eq11}) lead to the other Einstein equation
\be\label{eq15}
2\dot{H}+3H^2=-\frac{1}{f'(R)}\left[p_m+\ddot{R}f''(R)+\dot{R}^2f'''(R)+2H\dot{R}f''(R)+\frac{1}{2}\left(f(R)-Rf'(R)\right)\right]
\ee
which allows the identification of the effective curvature-pressure 
\be\label{eq16}
p_R=\frac{1}{f'(R)}\left[\ddot{R}f''(R)+\dot{R}^2f'''(R)+2H\dot{R}f''(R)+\frac{1}{2}\left(f(R)-Rf'(R)\right)\right]
\ee
Hence, the curvature terms in the r.h.s. of Eqs. (\ref{eq14}) and (\ref{eq16}) can be viewed as an effective
curvature-fluid with equation of state parameter expressed as 
\be\label{eq17}
\omega_R=-1+2\frac{\ddot{R}f''(R)+\dot{R}^2f'''(R)-H\dot{R}f''(R)}{Rf'(R)-f(R)-6H\dot{R}f''(R)}
\ee
defining the function $\Phi(R)=f'(R)$ \cite{sotiriou} and combining  Eqs. (\ref{eq13}) and (\ref{eq14}) it follows that
the EoS parameter for the curvature-fluid (\ref{eq17}) can be written as
\be\label{eq18}
\omega_R=-1+\frac{\ddot{\Phi}-H\dot{\Phi}}{3H^2\Phi-\rho_m}
\ee
where the last term $\rho_m$ is the pressureless dark matter contribution. 
According to the holographic correspondence this EoS parameter should be equal to the one given by Eq. (\ref{eq5}). This conduces to the following equation for the function $\Phi$
\be\label{eq19}
\ddot{\Phi}-H\dot{\Phi}-2\frac{\alpha-1}{\beta}\Phi H^2=-\frac{2}{3}\frac{\alpha-1}{\beta}\rho_m
\ee
It should be noted here that the cosmological evolution takes place in the Einstein frame as is clear from Eq. (\ref{eq11}), and therefore only in this frame the solution (\ref{eq4}) makes sense as this solution is obtained via solving the standard Einstein equations.\\
Changing the variable from cosmic time t to redshift z, the matter density $\rho_m$ becomes
\be\label{eq20}
\rho_m=3H_0^{2}\Omega_{m0}(1+z)^{3}
\ee
here we assumed that $\rho_m$ is conserved separately. After this change of variable, the equation (\ref{eq19}) transforms to
\be\label{eq21}
(1+z)^{2}H^{2}\frac{d^{2}\Phi}{dz^{2}}+(1+z)^{2}H\frac{dH}{dz}\frac{d\Phi}{dz}-2\frac{\alpha-1}{\beta}\Phi=
-2\frac{\alpha-1}{\beta}H_0^{2}\Omega_{m0}(1+z)^{3}
\ee
The general solution to this equation is given by 
\be\label{eq22}
\begin{aligned}
\Phi(z)=&\frac{2p\Omega_{m0}}{9p-6p^2-2}\left(1+z\right)^{3-2/p}+\Phi^{+}_0\left(1+z\right)^{(p-1+\sqrt{(p-1)^2+8p})/2p}\\
&+\Phi^{-}_0\left(1+z\right)^{(p-1-\sqrt{(p-1)^2+8p})/2p}
\end{aligned}
\ee
where $p=\frac{\beta}{\alpha-1}$ and the constants $\Phi^{\pm}_0$ will be determined using local solar system consistency.
To determine $f(R)$, we first reconstruct the form of $f(R)$ from (\ref{eq22}) as a function of the redshift z. From $\Phi(R)=f'(R)$ it follows that
\be\label{eq23}
\frac{df[R(Z)]}{dz}=\Phi(z)\frac{dR}{dz}
\ee
Expressing the scalar curvature $R=6(\dot{H}+2H^2)$ in terms of $z$ through the equation (\ref{eq20}) and replacing $\Phi$ from (\ref{eq22}), the Eq. (\ref{eq23}) can be integrated with respect to $z$, yielding
\be\label{eq24}
\begin{aligned}
f(z)=&C_1(p)H_0^2\Omega_{m0}(1+z)^3+C^{+}_2(p)H_0^2\Phi^{+}_0\left(1+z\right)^{(p+3+\sqrt{(p-1)^2+8p})/2p}\\
&+C^{-}_2(p)H_0^2\Phi^{-}_0\left(1+z\right)^{(p+3-\sqrt{(p-1)^2+8p})/2p}
\end{aligned}
\ee
where
\be
C_1(p)=\frac{8(1-2p)}{6p^3-9p^2+2p} , \,\, C^{\pm}_2(p)=\frac{24(2p-1)}{p(p+3\pm\sqrt{(p-1)^2+8p})}
\ee
writing the variable $(1+z)$ in terms of $R$ using Eq. (\ref{eq20}), we can obtain $f(R)$ explicitly in terms of $R$
\be\label{eq25}
\begin{aligned}
f(R)=&\left(\frac{p}{6H_0^2(2p-1)}\right)^{3p/2}H_0^2\Omega_{m0}C_1(p)R^{3p/2}\\
&+\left(\frac{p}{6H_0^2(2p-1)}\right)^{(p+3+\sqrt{(p-1)^2+8p})/4}H_0^2\Phi^{+}_0C^{+}_2(p)R^{(p+3+\sqrt{(p-1)^2+8p})/4}\\
&+\left(\frac{p}{6H_0^2(2p-1)}\right)^{(p+3-\sqrt{(p-1)^2+8p})/4}H_0^2\Phi^{-}_0C^{-}_2(p)R^{(p+3-\sqrt{(p-1)^2+8p})/4}
\end{aligned}
\ee
The constants of integration $\Phi^{\pm}_0$ can be determined using a local consistency conditions with the aim to not affect by measurement errors
the reconstructed $f(R)$. This conditions are $f'(R)_{z=0}=1$ (to recover the actual value of the Newtonian constant from Eq. (\ref{eq11}) at $z=0$) and $f''(R)_{z=0}=0$ (consistency with the weak field approximation), which translates into (see \cite{rainer},\cite{cardone})
\be\label{eq26}
\left(\frac{df}{dz}\right)_{z=0}=\left(\frac{dR}{dz}\right)_{z=0}, \,\,\left(\frac{d^2f}{dz^2}\right)_{z=0}=\left(\frac{d^2R}{dz^2}\right)_{z=0}
\ee
then, using this conditions the constants $\Phi^{\pm}_0$ become
\be\label{eq27}
\begin{aligned}
\Phi^{\pm}_0=&3\Big[\mp8(2p^2-5p+2)\mp4(2p-1)(3-p\mp\sqrt{p^2+6p+1})\\
&+p^2C_1(p)\Omega_{m0}(3-5p\mp\sqrt{p^2+6p+1})\Big]/\\
&\left[pC^{\pm}_2(p)\sqrt{p^2+6p+1}(p+3\pm\sqrt{p^2+6p+1})\right]
\end{aligned}
\ee
Note that this reconstruction was made in the low redshift region, where the power law expansion resulting from the restriction of the holographic model to the dark energy dominance (see (\ref{eq3})), is an acceptable approximation according to the data, as shown in Fig. 1. Note also that our model is approximate as is obtained from reconstruction
scheme and it is valid within the adopted approximation. It is not so easy to recover more realistic non-linear f(R) gravity which gives theory \ref{eq25} as the approximation.

\section{Discussion}
In this work we showed a reconstruction of the late time cosmological $f(R)$ gravity dynamics, based on the holographic energy density cut-off proposed in \cite{granda},\cite{granda1}. An analytical expression for the $f(R)$ model compatible with observational data at low redshift has been found.
The reconstructed Lagrangian is of a power-law type and contains only positive curvature powers for $\alpha>1, \beta>0$ (quintessence phase) and negative curvature powers for $\alpha<1, \beta>0$ (phantom phase), except in the region of parameters $-3-\sqrt{8}<\beta/(\alpha-1)<\sqrt{8-3}$, which is forbidden in this model as the power becomes complex. The $\alpha<1, \beta<0$ can also be considered, giving similar to the case $\alpha>1, \beta>0$ results. From equations (\ref{eq3}) and (\ref{eq5}) it follows that for $\alpha=1$ we obtain the de Sitter solution with $\omega_{\Lambda}=-1$. The last two terms in (\ref{eq25}) correspond to the homogeneous solution of (\ref{eq19}) with $\Omega_{m0}=0$, which describe a purely geometrical-dominance model of dark energy . For the specific case of $\alpha=1.1, \beta=0.5$ we expect that the term with lower power $R^{2-\sqrt{7}/4}$ dominates. The matter dominance epoch occurs for $p=2/3$, giving the Einstein term plus corrections (this corrections may be caused due to the extrapolation to high redshift).
Is important to note that the holographic dark energy density used here was obtained within the Einstein frame and not as the solution to the modified $f(R)$ gravity. Hence the reconstructed $f(R)$ theory effectively describes the holographic dark energy in Einstein gravity. Nevertheless, Fig. 1 shows that the proposed solution (\ref{eq4}) is a good starting point, at least at the phenomenological level, to reconstruct  the $f(R)$ model at low redshift independent of the theory.\\
In summary, we have studied how the modified $f(R)$ gravity  model can be
used to describe the holographic energy density as an effective theory at low redshift, which conduces to the reconstruction in a direct and unambiguous way. However, this reconstruction is settled at the phenomenological level, and the theoretical root of the holographic density still to be investigated.
\section*{Acknowledgments}
This work was supported by the Universidad del Valle.


\begin{thebibliography}{99}
\bibitem{tegmark} M. Tegmark et al. [SDSS Collaboration], Phys. Rev. D 69, 103501 (2004) [astro-ph/0310723];
K. Abazajian et al. [SDSS Collaboration], Astron. J. 128, 502 (2004) [astro-ph/0403325];
K. Abazajian et al. [SDSS Collaboration], Astron. J. 129, 1755 (2005) [astro-ph/0410239].
\bibitem{riess} A.G. Riess, et al., Astron. J. \textbf{116}, 1009 (1998), [astro-ph/9805201]; Astron. J. \textbf{117},707 (1999);
S.Perlmutter \textit{et al}, Nature \textbf{391}, 51 (1998); S. Perlmutter et al. [Supernova Cosmology Project Collaboration], Astrophys. J. 517, 565 (1999) [astro-ph/9812133]; P. Astier et al., Astron. Astrophys. \textbf{447}, 31 (2006) [astro-ph/0510447].
\bibitem{spergel}D. N. Spergel et al. [WMAP Collaboration], Astrophys.
J. Suppl. 148, 175 (2003) [astro-ph/0302209]; D. N. Spergel et al., astro-ph/0603449.
\bibitem{copeland} Edmund J. Copeland, M. Sami and Shinji Tsujikawa, Int. J. Mod. Phys. D \textbf{15}
1753-1936 (2006), arXiv:hep-th/0603057
\bibitem{padmana} T. Padmanabhan, Physics Reports \textbf{380} (2003) 235–320, hep-th/0212290
\bibitem{sergei1} S. Nojiri and S.D. Odintsov,Phys. Lett. \textbf{B576} (2003) 5, hep-th/0307071
\bibitem{sergei2} S. Nojiri and S.D. Odintsov, Phys.Rev. \textbf{D68} (2003)123512, hep-th/0307288
\bibitem{sergei3} S. Nojiri and S.D. Odintsov, Gen. Rel. Grav. \textbf{36} (2004) 1765, hep-th/0308176
\bibitem{carroll} S. M. Carroll, V. Duvvuri, M. Trodden and
M. S. Turner, Phys. Rev. D \textbf{70}, 043528 (2004), arXiv:astro-ph/0306438
\bibitem{capozziello} Salvatore Capozziello, Sante Carloni and Antonio Troisi, Recent Res. Dev. Astron. Astrophys 1 (2003), 625, astro-ph/0303041
\bibitem{sergei4} M.C.B. Abdalla, S. Nojiri and S.D. Odintsov, Gen. Rel. Grav. \textbf{36} (2004) 1765, hep-th/0308176
\bibitem{sergei41} S. Nojiri and S.D. Odintsov, Phys. Rev. \textbf{D77} (2008),026007, hep-th/0710.1738
\bibitem{sergeinoj} S. Nojiri and S.D. Odintsov, hep-th/0807.0685
\bibitem{sergei5} S. Nojiri and S.D. Odintsov, Phys. Lett. \textbf{B631}, 1, 2005 , hep-th/0508049
\bibitem{sergei6} Guido Cognola, Emilio Elizalde, Shin'ichi Nojiri, Sergei D. Odintsov and Sergio Zerbini, Phys. Rev. \textbf{D73}, 084007, 2006, hep-th/0601008
\bibitem{sergei7} Guido Cognola, Emilio Elizalde, Shin'ichi Nojiri, Sergei D. Odintsov and Sergio Zerbini,JCAP, 0502:010, 2005, hep-th/0501096
\bibitem{sergei8} S. Nojiri and S.D. Odintsov, Phys. Rev. \textbf{D74} (2006), 086005, hep-th/0608008
\bibitem{sergei9} S. Nojiri and S.D. Odintsov, Int. J. Geom. Meth. Mod. Phys. 4, 115, 2007, hep-th/0601213
\bibitem{sergei10} S. Nojiri and S.D. Odintsov, Phys. Lett. \textbf{B652}, (2007), 343, hep-th/0706.1378
\bibitem{sergei11} S. Nojiri and S.D. Odintsov, Phys. Lett. \textbf{B657}, (2007), 238, hep-th/0707.1941    
\bibitem{sergei12} S. Capozziello, S. Nojiri and S.D. Odintsov and A. Troisi, Phys. Lett. \textbf{B639}, (2006), 135, astro-ph/0604431
\bibitem{sergei13} S. Nojiri and S.D. Odintsov, J. Phys. \textbf{A40}, (2007), 6725, hep-th/0610164
\bibitem{sergei14} S. Nojiri and S.D. Odintsov, J. Phys. Conf. Ser. 66 (2007), 012005, hep-th/0611071
\bibitem{sergei15} F. Briscese, E. Elizalde, S. Nojiri and S. D. Odintsov, Phys. Lett. \textbf{B646}, (2007),105, hep-th/0612220
\bibitem{sergei16} S. Nojiri, S.D. Odintsov and Petr V. Tretyakov, Phys. Lett. \textbf{B651} (2007), 224, hep-th/0704.2520
\bibitem{cardone} S. Capozziello, V. F. Cardone and A. Troisi, Phys. Rev. \textbf{D71}, 043503 (2005)
\bibitem{xing} Xing Wu and Zong-Hong Zhu, Phys. Lett. \textbf{B660} (2008), 293, astro-ph/0712.3603
\bibitem{sergei17} S. Nojiri and S.D. Odintsov, Phys. Lett. \textbf{B659}, (2008), 821, hep-th/0708.0924  
\bibitem{sergei18} S. Nojiri, S.D. Odintsov and Petr V. Tretyakov, Prog. Theor. Phys. Suppl. 172 (2008),81, hep-th/0710.5232  
\bibitem{sergei19} Kazuharu Bamba and S.D. Odintsov, Phys. Lett. \textbf{B659}, (2008), 821, hep-th/0801.0954 
\bibitem{sergei20} Kazuharu Bamba, S. Nojiri and S.D. Odintsov, Phys. Rev. \textbf{D77} (2008), 123532, hep-th/0803.3384
\bibitem{sergei21} S. Nojiri and S.D. Odintsov, Phys. Rev. \textbf{D78}, (2008), 046006, hep-th/0804.3519  
\bibitem{sergei22} Kazuharu Bamba, S. Nojiri and S.D. Odintsov, JCAP0, 810, 045 (2008), hep-th/0807.2575 
\bibitem{sergei23} S. Capozziello, M. De Laurentis, S. Nojiri and S.D. Odintsov, hep-th/0708.1335 
\bibitem{li} M. Li, Phys. Lett. B \textbf{603}, 1 (2004), [hep-th/0403127]
\bibitem{granda} L.N. Granda and A. Oliveros, Phys. Lett. B \textbf{669}, 275-277 (2008), arXiv:gr-qc/0810.3149
\bibitem{sergei24} S. Nojiri and S.D. Odintsov, Gen. Rel. Grav. \textbf{38} (2006) 1285, hep-th/0506212
\bibitem{granda1} L. N. Granda and A. Oliveros, gr-qc/0810.3663
\bibitem{padmanabhan} T. Roy Choudhury and T. Padmanabhan,  Astron. Astrophys. 429 (2005), 807, astro-ph/0311622
\bibitem{riess1} Riess A. G., Strolger L., Tonry J., et al., ApJ 607 (2004), 665
\bibitem{sotiriou} Thomas P. Sotiriou and Valerio Faraoni, gr-qc/0805.1726
\bibitem{rainer}  Rainer Dick, Gen. Rel. Grav. 36, 82004),217, gr-qc/0307052

\end{thebibliography}
\end{document}